\documentstyle[epsf,aps,twocolumn]{revtex}

\begin{document}
\draft
\sloppy
\wideabs{
\title{Density-functional study of hydrogen chemisorption on vicinal 
       Si(001) surfaces} 
\author{E.~Pehlke}
\address{Physik Department T30, Technische Universit\"at M\"unchen, 
         D-85747 Garching, Germany}
\author{P. ~Kratzer}
\address{Fritz-Haber-Institut der Max-Planck-Gesellschaft, Faradayweg 4--6,
         D-14195 Berlin-Dahlem, Germany}  
\date{\today}
\maketitle

\begin{abstract}
Relaxed  atomic geometries and chemisorption energies have been
calculated for the dissociative adsorption of 
molecular hydrogen on vicinal Si(001) surfaces. 
We employ density-functional theory, 
together with a pseudopotential for Si, and 
apply the generalized gradient approximation by Perdew and Wang
to the exchange-correlation functional.
We find the double-atomic-height rebonded D$_{\rm B}$ step, 
which is known to be stable on 
the clean surface, to remain stable on partially hydrogen-covered surfaces. 
The H atoms preferentially bind to the Si atoms at the rebonded step edge, 
with a chemisorption energy difference with respect to the terrace sites 
of ${_> \atop ^\sim}0.1$~eV. A surface with rebonded single 
atomic height S$_{\rm A}$ and  S$_{\rm B}$ steps gives very similar results.
The interaction between H--Si--Si--H mono-hydride units is shown to be 
unimportant for the calculation of the step-edge hydrogen-occupation.
Our results confirm the interpretation and results of the recent 
H$_2$ adsorption experiments on vicinal Si surfaces 
by Raschke and H\"ofer described in the preceding paper. 
\end{abstract}
}
\section{Introduction}
\label{sec:introduction}

The interaction of hydrogen with silicon surfaces has become an intensively
studied matter. There are important applications in semiconductor 
technology, such as the passivation of surfaces, etching, and 
chemical vapor deposition growth.\cite{lagally:95,owen:97a,owen:97b} 
The attachment of the deposited Si atoms at
surface steps is an essential  aspect of epitaxial 
growth.\cite{hamers:89a,mo:90a,vvedensky:90,lagally:93,lagally:97} 
From the difference in hydrogen chemisorption energy between the adsorption
sites on the terraces and at the steps of a vicinal Si(001) surface
the equilibrium hydrogen occupation of the various surface 
sites can be derived. 
If hydrogen atoms preferentially bind to the step edge, already
small hydrogen coverages are sufficient to saturate the Si dangling
bonds at the step edge, and thereby to affect the probability of
Si atoms to become attached to the step. In this way the kinetics of
Si epitaxial growth can be affected by the presence of hydrogen. 
Of course there are also other mechanisms, like the hydrogen-induced 
change of the Si surface diffusion coefficient.\cite{lagally:95}
Furthermore, within the framework of thermodynamic equilibrium theory, the
step energies govern the surface morphology on a mesoscopic 
lengthscale.\cite{alerhand:88,alerhand:90,tromp:92}
As step energies depend on hydrogen coverage, the adsorption
of hydrogen on the surface could affect step roughness
or even surface morphology.\cite{hoegen:96}

Last but not least, the dissociative adsorption and associative desorption of 
molecular hydrogen on the Si(001) surface has attracted a great attention,
in particular because adsorption and
desorption experiments have led to apparently contradictory results
with respect to the adsorption energy barrier.
On the one hand, the observed small sticking coefficient suggests a
large adsorption energy barrier, while, on the other hand, the kinetic
energy distribution of the desorbing hydrogen molecules is nearly
thermal, suggesting only a small adsorption energy 
barrier.\cite{kolasinski:94,bratu:96a}
Despite intensive research there are still competing explanations.
In the dynamical model by Brenig {\it et al.}\cite{brenig:94a,bratu:96b} 
the desorption proceeds
from two hydrogen atoms bound to the two Si atoms of a single  
surface dimer. This hydrogen pairing is corroborated by the observed
first-order desorption kinetics of the hydrogen, and the deviations 
towards second order at very small hydrogen coverage.\cite{hoefer:92} 
Brenig's model contains a barrier; the small kinetic energy of the desorbing
particles is ascribed to an efficient energy transfer into the Si
surface phonon degrees of freedom during the desorption process.
However, several quantum-chemical 
cluster-calculations\cite{carter:91,carter:93,whitten:93a,nachtigall:96} 
have arrived at distinctly
larger barriers than density-functional 
slab-calculations,\cite{kratzer:94,pehlke:95a,vittadini:95a} 
and these large desorption energy barriers 
appear to be at variance with the desorption dynamics sketched above.
Thus the authors of these calculations favor models, in which the
adsorption on and desorption from surface defects play a 
major role.\cite{whitten:93b,nachtigall:94a,carter:96a,carter:97}
On the other hand, Pai and Doren\cite{doren:95} obtained barriers from 
density-functional cluster calculations that are consistent with the
pre-pairing model.

To investigate the role of steps and defects experimentally, Raschke and 
H\"ofer\cite{raschke:98a}
have studied the adsorption of H$_2$ on vicinal Si(001) surfaces. 
For studied miscut angles larger than 2$^\circ$ double atomic
height D$_{\rm B}$ steps prevail on the 
clean surface.\cite{alerhand:90,umbach:91a}
Raschke and H\"ofer infer from their experimental results 
that H$_2$ molecules preferentially adsorb at the
step sites. Most importantly, they can distinguish between the 
contribution of terraces and steps to the H$_2$-sticking coefficient. 
Contrary to the large barrier towards H$_2$ adsorption on 
the flat surface and on the terraces, 
there appears to be only a rather minor ($\sim$0.09 eV) adsorption 
energy barrier along the reaction path that leads 
to dissociative adsorption at the step edge.\cite{raschke:98b}
The measured hydrogen step-saturation coverage of about one 
H atom per (1$\times$1) surface lattice constant $a$ along the step
is consistent with a model in which the hydrogen atoms
bind to the edge of the double atomic height steps in a local 
mono-hydride like geometry. 
Furthermore, at elevated temperatures Raschke and H\"ofer observe the 
diffusion of the adsorbed hydrogen atoms from the step onto the terrace. From
the measured equilibrium hydrogen coverage on the step edge 
and on the terrace 
they deduced a difference between the local chemisorption energies 
of roughly 0.15 -- 0.3 eV by which the hydrogen atoms bind more strongly to the 
step edge Si atoms.\cite{raschke:98c}

In view of the great importance of H$_2$/Si(001) as a model system 
to understand the dynamics of adsorption and desorption, and in view of
the conflicting results of quantum-chemical and density-functional 
theory (DFT) based computations of the adsorption energy barrier 
of the flat surface,
it appears to be quite desirable to calculate the H$_2$
chemisorption energies and adsorption energy barriers for 
vicinal surfaces.
A successful comparison of DFT results to the new experimental data 
would lend support to both the interpretation of the experiment and
the credibility of the generalized gradient approximation to
the exchange-correlation functional used in the DFT calculations.

The purpose of this work is to provide a comprehensive overview 
over H$_2$ chemisorption on
Si(001) vicinal surfaces: Relaxed geometries and chemisorption 
energies will be presented for various adsorption sites and step
topologies. The adsorption energy barriers on the terrace and
at the step edge will be discussed in a forthcoming paper.\cite{raschke:98b}

The organization of the paper is as follows:
First, the stable configurations of the clean steps (which are a function
of miscut angle) are reviewed in Section \ref{sec:clean}. 
These structures have already been thoroughly investigated 
both experimentally\cite{wierenga:87,wierenga:89,swartzentruber:90a} and 
theoretically.\cite{chadi:87,boguslawski:94,oshiyama:95}
However, the stability of these steps in the presence of hydrogen is not 
{\it a priori} obvious, and to our knowledge has not been 
calculated before.
Therefore hydrogen adsorption on three different atomic step 
topologies, which have previously been 
discussed in context with the clean surface,\cite{chadi:87} 
will be considered in Sections \ref{sec:sasb} -- \ref{sec:nonrebdb}. 
A comparison between these results can lead to a better understanding
of the mechanism leading to the different chemisorption energies.
Finally, the interaction between neighboring mono-hydride 
groups on the surface will be discussed in
Section \ref{sec:hcov}. A quantitative comparison to 
Raschke and H\"ofer's\cite{raschke:98c}
experimental data is given in Section \ref{sec:lowcov}.

\section{Computational method}
\label{sec:computational}

The total-energy minimizations and geometry optimizations have been 
carried out using the electronic-structure code {\tt fhi96md}\cite{fhi96md}
in a version parallelized for the CRAY T3E architecture. 
In this code total energies and forces acting on the atoms are 
calculated within DFT.
The generalized gradient approximation (GGA) by
Perdew and Wang\cite{perdew:92} is applied to the 
exchange-correlation energy functional.
The GGA is known to be distinctly more reliable than the local
density approximation (LDA) especially with respect to binding
energies, while the LDA has proven insufficient for the calculation
of energy barriers of reactions involving 
hydrogen.\cite{perdew:92,hammer:94,moll:95} Though
recently Nachtigall and Jordan\cite{nachtigall:96} 
have argued that even the GGA is not
sufficiently accurate for the calculation of energy barriers, 
the use of the GGA is expected to be fully adequate for the
less demanding problem of
calculating energy differences for hydrogen atoms adsorbed 
on - in our case even chemically rather similar - sites on the 
Si surface.

The Si atoms are represented by pseudopotentials, which have
been constructed according to Hamann's scheme,\cite{hamann:89,fuchs:98b} 
consistently using the same GGA for the construction of the pseudopotential
and the solid state calculations.\cite{fuchs:98a} For the hydrogen atoms we
take the full $1/r$ Coulomb potential.

The electronic wave functions are expanded into plane waves 
up to a cut-off energy $E_{\rm cut}$. 
The integration over the Brillouin zone is replaced by a summation
over one or two special {\bf k} points.\cite{monkhorst:76} 
The {\bf k} points  are chosen equidistant and
aligned along the direction of the step edge. We assume no symmetry restrictions
to the atomic geometry during relaxation, the only symmetry 
exploited in these calculations is time reversal.
To give the reader an impression of the residual error due to the 
finite cut-off energy and {\bf k}-point set, we will quote below
numerical results at two levels of accuracy: 
The (30 Ry, 1 {\bf k}) data result from geometry
relaxation runs using $E_{\rm cut}=$30 Ry (408 eV) plane-wave cut-off energy 
and one special {\bf k} point (not the $\Gamma$ point). 
Important calculations have been repeated with 50 Ry (680 eV) cut-off energy
and 2 {\bf k} points, using the frozen geometry from the
previous (30 Ry, 1 {\bf k}) run.  
At 50 Ry cut-off energy there are about 153000 complex plane wave 
coefficients for every band and {\bf k}-point.

Total energies are calculated for Si(1~1~11) slabs
with a thickness of about six atomic layers, using a supercell
that is periodically repeated in all three dimensions.
This surface orientation corresponds to a miscut angle of 
7.3$^\circ$ in the [110] direction away from (001). 
On the D$_{\rm B}$ stepped 
surface the (1$\times$2) dimerized terraces are four dimers
wide, ensuring that we can neglect elastic 
step-step interactions\cite{poon:92}  
when generalizing the results to smaller miscut angles.
Furthermore, the large terrace width allows us to distinguish
between dimers close to the step edge and those 
in the center of the terrace.
The width of the supercell amounts to $2a$, allowing for
one Si-dimer row perpendicular to the step edge.  
The surface unit cell on the top of the slab contains either 
a single double atomic height 
step or a pair of single atomic height steps. 
The atoms in the bottom two Si layers are fixed at their bulk
positions. We use the theoretical equilibrium lattice
constant $c = a \sqrt{2} = 5.450$ \AA. On the bottom surface
the slab is saturated with H atoms in a local di-hydride configuration. 
In case of the clean surface our supercell
contains altogether 74 Si atoms and 22 H atoms.
All atoms apart from those in the bottom two Si layers 
and the bottom H-termination are allowed to relax 
by following the computed Hellman-Feynman forces. 
Residual forces are smaller than $3\times 10^{-2}$ eV/\AA.

All results quoted below are plain total-energy differences, not 
corrected for the zero-point energy (ZPE) of the atoms.
For the vibration of the free H$_2$ molecule the ZPE 
is $\hbar \omega/2 = 0.272$ eV.
For hydrogen atoms adsorbed on a Si surface in a mono-hydride like
configuration the frequency of the Si--H vibrations does not
depend much on the details of the configuration.\cite{stucki:83} 
Results from electron energy loss spectroscopy indicate for
Si(001) (2$\times$1) H one stretching mode at 0.260 eV,
and two bending modes at 0.0785 eV.\cite{stucki:83} For two adsorbed
H atoms this gives a ZPE of 0.417 eV. We neglect changes of the Si 
phonon frequencies due to hydrogen adsorption. 
Within this approximation the ZPE correction leads to only a rigid
shift of all chemisorption energies by 0.145 eV.
For deuterium this shift would be smaller by about a factor 
of $1/\sqrt{2}$.  Differences between chemisorption energies are not 
affected by the ZPE correction, i.e., they can be directly read
from the tables below.

\section{Results and discussion}
\label{sec:results}

In the following we present our DFT results and 
discuss the implications for the thermodynamics of partially 
H-covered vicinal Si surfaces. 
First, however, we briefly summarize our results for the step
energies and relaxed geometries of the clean surface.
In the subsequent Sections \ref{sec:sasb} -- \ref{sec:nonrebdb}
the chemisorption energies will always refer to the respective
clean surface plus a hydrogen molecule at rest far away from the 
surface as energy zero. Thus, in order to find out which step 
topology is stable (at zero temperature), one first has to add 
the respective step energies from Section \ref{sec:clean} to 
the chemisorption energies before comparing the total energies of
different step topologies with each other.
As explained in the previous Section \ref{sec:computational},
we expect the contributions from the zero point energy to 
cancel out when a total energy difference between 
rather similar, mono-hydride like configurations
is calculated.

It should be noted here that the probability of observing
a certain type of step (e.g., S$_{\rm A}$-S$_{\rm B}$ or
D$_{\rm B}$) along a step edge at some finite
temperature cannot be simply inferred from the respective step
energies per unit step length. The step energy difference per 
unit length has to be multiplied with an appropriate 
coherence length first (which describes the average extent
of a kink-free part of the step) in order to obtain an excitation energy
that makes sense to enter in a Boltzmann factor.\cite{kochanski:90}

Our objective is to characterize the lowest energy configurations
for two H atoms adsorbed on the vicinal Si surface. 
The pairing energy for two hydrogen atoms on the Si(001) surface
has been measured by H\"ofer, Li, and Heinz\cite{hoefer:92} 
to be about 0.25 eV, 
which is close to the DFT result by Northrup.\cite{northrup:95} 
This value is
larger than the chemisorption energy difference between typical 
adsorption sites. Therefore we have restricted ourselves to 
configurations with two hydrogen atoms adsorbed on the same 
silicon dimer, i.e., we have only considered the H-paired 
configurations.
Of course the unpaired H-configurations will become important
when comparing to an experiment carried out at high temperature
and low H-coverage, where entropy plays an essential role. 
We postpone this issue until Section \ref{sec:lowcov}.

\subsection{Atomic geometry of clean vicinal Si(001) surfaces}
\label{sec:clean}

A characteristic  feature of the flat Si(001) surface are the 
Si dimers.\cite{farnsworth:59} 
By forming these dimers the density of dangling bonds 
is halved in comparison to the bulk terminated surface. 
The Si dimers are not parallel to the 
surface.\cite{chadi:79,dabrowski:92,wolkow:92,shkrebtii:93,bullock:95,over:97} 
In our calculation the buckling angle amounts to about 19$^\circ$ 
on the p(2$\times$2) dimerized (001) terrace, 
in agreement with experiment\cite{bullock:95,over:97} 
and previous DFT work.\cite{northrup:93a,ramstad:95}
The energy lowering is driven by the re-hybridization of the $sp^3$
orbitals of the Si atom that relaxes towards the bulk:
Its bonding geometry becomes more planar, 
and therefore the bonding orbitals become more $sp^2$ like,
while the dangling bond gains more $p$ character and becomes unoccupied.
The dangling bond of the other Si dimer atom that relaxes outwards
is fully occupied.
This implies that the $\pi$ bond between the Si atoms 
of the symmetric dimer is partially destroyed by buckling, however,
in case of Si the energy gain due to the re-hybridization
mechanism obviously over-compensates this energy cost.
The buckling of the surface dimers is accompanied by
a change from a metallic to a semiconducting electronic
surface band structure.\cite{pollmann:95a}
Due to elastical coupling via the atoms in the second and deeper
layers the buckling angle alternates along the dimer row.
The lowest energy reconstruction is p(2$\times$2),
or, even slightly lower, c(4$\times$2), but the 
energy difference between these reconstructions is so small 
that it is irrelevant for our purpose.\cite{ramstad:95} 
Thus we assume the p(2$\times$2) reconstruction on
the terraces, which allows us to keep the supercell small.
 
\begin{figure}[htb]
  \epsfxsize=6.0cm
  \centerline{\epsfbox{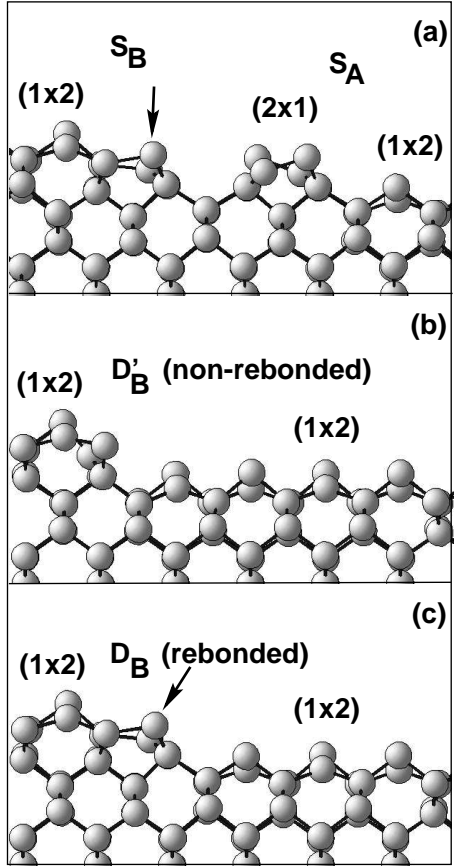}}
  \caption{Relaxed atomic geometry of single and double 
           atomic height steps on the Si(001) surface.
	   The orientation of the vicinal surface 
	   is (1~1~11), corresponding to a miscut angle 
	   of $\Theta = 7.3^\circ$. 
	   A side view along the [1\=10] direction
	   parallel to the step edge is shown.
           (a) Pair of single atomic height S$_{\rm A}$
	   and S$_{\rm B}$ steps. Terraces are 
           alternate (1$\times$2) and (2$\times$1)
	   dimerized.
	   (b) Non rebonded double atomic height
           D$_{\rm B}^\prime$ step.  
	   (c) Rebonded double atomic height D$_{\rm B}$
	   step.
	   In panels (a) and (c) the rebonded Si atoms 
	   at the S$_{\rm B}$ and D$_{\rm B}$ step edge
	   are denoted by arrows.
          }
  \label{fig:clean_surfaces}
\end{figure}

The vicinal Si(001) surface can be imagined as a staircase with
(001) terraces separated by steps.
At small miscut angles less than $\sim 1^\circ$, the vicinal surface 
consists of alternating single atomic height S$_{\rm A}$ 
and S$_{\rm B}$ steps,\cite{alerhand:90}
and, consequently, the direction of dimerization
rotates by 90$^\circ$ on successive terraces. 
There occurs a presumably gradual transition initiating at about $1^\circ$ to
$2^\circ$, driven by elastic step-step interactions, and at 
large miscut angles the surface 
becomes more and more single-domain with predominantly double atomic height 
steps.\cite{alerhand:90,bennett:91,pehlke:91b,umbach:91a,miguel:92}
These steps have been identified both experimentally and
theoretically to display the rebonded D$_{\rm B}$ geometry, with threefold
coordinated Si atoms at the step 
edge.\cite{chadi:87,wierenga:87,wierenga:89}

We use the clean Si surface as the energy reference for the chemisorption 
energies discussed below. Therefore we have computed the relaxed geometries 
and total energies for the three step configurations displayed in 
Fig.~\ref{fig:clean_surfaces}.
Our results turned out to be in agreement with previous DFT studies
of steps on the Si(001) surface.\cite{boguslawski:94,oshiyama:95}
However, note that the choice of, e.g., the plane-waves cut-off energy 
is determined by the requirement to accurately describe hydrogen-containing 
systems, it is not adapted to the calculation of Si step energies, for
which a much smaller cut-off would be sufficient. However,
to obtain accurate step interaction energies , thicker Si slabs
would be necessary.\cite{poon:92}  

An S$_{\rm A}$ - S$_{\rm B}$ step pair is displayed in 
Fig.~\ref{fig:clean_surfaces}(a). The (1$\times$2) and (2$\times$1) 
terraces are 
both $2a$ wide. The bonds between the two rebonded Si atoms and the 
neighboring step-edge atom are highly strained, they are 4\% and 7\% 
longer than the bulk Si--Si bond.
The Si surface dimer bonds are 0 -- 2.4\% shorter than the bulk bond length,
and the buckling angle varies between 18$^\circ$ and 15$^\circ$, i.e., 
the buckling is partially suppressed close to the steps.
The electronic structure is characterized by occupied dangling bonds at the
Si dimer and step edge atoms that have relaxed outwards.
The re-hybridization mechanism works for the rebonded Si atom at the step edge
in the same way as for the Si dimer atoms, it results in a height difference 
between neighboring rebonded step edge atoms of 0.7 \AA.
This is close to the result by Bogus{\l}awski {\it et al.}\cite{boguslawski:94}
of 0.58 \AA.

The double atomic height non-rebonded D$_{\rm B}^\prime$ and rebonded 
D$_{\rm B}$ steps are shown in Fig.~\ref{fig:clean_surfaces}(b) and (c),
respectively. 
The non-rebonded geometry corresponds to a double layer of Si, which is 
terminated at the step edge.  There are no rebonded Si atoms with highly 
strained bonds, 
thus the elastic interaction of the D$_{\rm B}^\prime$ steps is small. 
Among the step structures considered, this one has the
smallest strain field on the terraces. Therefore the dimers in the
middle of the (1$\times$2) terrace between the D$_{\rm B}^\prime$ 
steps most closely resemble the dimers on the flat surface.
The buckling angle of the surface dimers varies between 18$^\circ$
close to the step and 19$^\circ$ in the middle of the terrace. 
Again, re-hybridization results in a pronounced buckling at the step edge. 
However, there is an important difference between the rebonded 
configurations (a) and (c) and the non-rebonded D$_{\rm B}^\prime$ step: 
We find the Si--Si bond between the two Si atoms on the upper
terrace closest to the step edge to be contracted by 3\%,
which indicates a strong additional $\pi$-bond between these 
atoms.\cite{oshiyama:95}

It was already pointed out by Chadi\cite{chadi:87} 
that the non-rebonded D$_{\rm B}^\prime$
step is not the lowest energy configuration. The number of dangling
bonds can be lowered by adding an extra ``rebonded'' Si atom to 
the D$_{\rm B}^\prime$ step edge. The rebonded D$_{\rm B}$ 
structure is shown in 
Fig.~\ref{fig:clean_surfaces}(c), it can be described as a collapsed 
S$_{\rm A}$ - S$_{\rm B}$ step pair, with the width of the (2$\times$1) 
terrace shrunk to zero.
Similar to the S$_{\rm B}$ step, the bonds between the two rebonded Si atoms
and the neighboring step-edge atoms are highly strained, 4\% and 7.7\%
longer than a Si--Si bulk bond 
(4.7\% $\sim$ 6.6\% in Ref. \onlinecite{oshiyama:95}).
The force dipole from these bonds induces a strain field and results in a
considerable step-step interaction.
We find a variation of the dimer buckling angle between 18$^\circ$ in
the middle of the terrace and 16$^\circ$ on the lower terrace 
close to the step edge. The dimer bond lengths vary between 
0\% and 2\% contraction.
Compared to the S$_{\rm A}$ - S$_{\rm B}$ step pair we find the 
D$_{\rm B}$ step to be $\sim$0.02 eV per step-length $2a$ lower in 
energy. However, note that this energy difference is smaller than
our rough error estimate of $\sim$0.05 eV, thus the agreement 
with observation is partially fortuitous.

To compare the energy of the rebonded and non-rebonded double atomic
height steps we have calculated the Si chemical potential by using the
same supercell and convergence parameters as above and adding 
four Si atoms to the bulk of the slab while keeping the top and
bottom surface structures intact.
In agreement with previous work we find the non-rebonded D$_{\rm B}^\prime$
step to be energetically unfavorable. However, our energy difference
of 0.13 eV/$a$ is larger than Oshiyama's\cite{oshiyama:95} 
result of 0.06 eV/$a$.
This may be connected to the fact that we observe a more pronounced
relaxation of the atoms at the D$_{\rm B}$ step edge, a
height difference of the rebonded atoms of 0.6 \AA{} in contrast to
the much smaller value of 0.17 \AA{} in Ref. \onlinecite{oshiyama:95}.
We think that our large step-edge relaxation is corroborated
by the similarity between the D$_{\rm B}$ and S$_{\rm B}$ steps 
together with 
the fact that we agree with Bogus{\l}awski {\it et al.}\cite{boguslawski:94} 
with respect to the pronounced relaxation of the S$_{\rm B}$ step.

In view of the smaller number of dangling bonds at the D$_{\rm B}$
step the energy lowering with respect to the D$_{\rm B}^\prime$ step
appears to be rather small. This can be explained by the extra $\pi$-bonds
stabilizing the D$_{\rm B}^\prime$ step, and the large Si--Si bond strain  
destabilizing the D$_{\rm B}$ step. Both effects diminish the 
energy difference suggested by simple bond-counting.

\subsection{Hydrogen adsorbed on surfaces with S$_{\rm A}$-S$_{\rm B}$ steps}
\label{sec:sasb}

\begin{figure}[htb]
  \epsfxsize=8.7cm
  \epsfbox{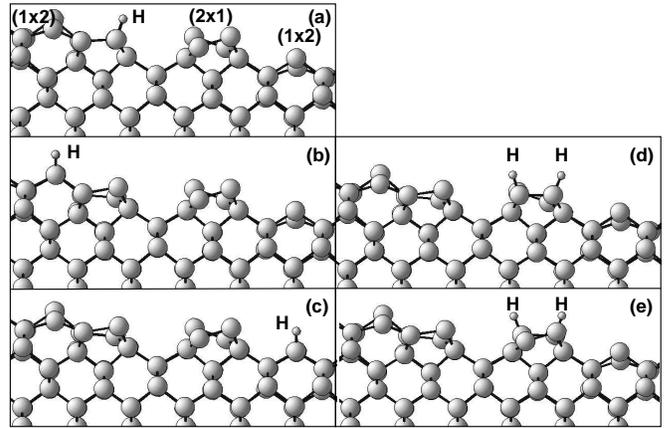}
  \caption{Relaxed atomic geometries for two hydrogen atoms adsorbed at
           different sites on a Si(1~1~11) surface with single atomic
	   height steps. 
	   Side view along the [1\=10] direction as in 
           Fig.~\ref{fig:clean_surfaces}. In panels (a) -- (c)
           the hydrogen atoms are arranged in a row 
           parallel to [1\=10], thus only the atom in the front 
	   is visible.
	   (a) Two hydrogen atoms adsorbed at the S$_{\rm B}$ 
           step edge in a local mono-hydride configuration.
           (b) Two hydrogen atoms adsorbed at the Si dimer on 
           the upper (1$\times$2) terrace 
           close to the S$_{\rm B}$ step. (c) Two hydrogen atoms
	   adsorbed at the Si dimer on the lower (1$\times$2) 
           terrace close to the S$_{\rm A}$ step. 
	   (d) and (e) Two hydrogen atoms adsorbed at a dimer on
           the (2$\times$1) terrace. 
          }
  \label{fig:h2onsasb}
\end{figure}

The relaxed geometries for two hydrogen atoms adsorbed on a 
surface dimer or at the two rebonded Si atoms are shown in
Fig.~\ref{fig:h2onsasb} for a vicinal surface with 
single atomic height steps. 
In panel (a) the two H atoms saturate the dangling bonds of the 
rebonded Si atom at the S$_{\rm B}$ step edge. Therefore there is no
driving force towards buckling anymore and the two rebonded Si atoms
become nearly equivalent. The Si--Si bond between the Si step edge
atom and its neighbor on the upper terrace is strained by about 5\%
compared to the Si--Si bulk bond length. This is almost identical 
to the average strain of these bonds on the clean surface, i.e.,
the step interaction will not be much affected by adsorption.

On the whole, the geometry changes upon hydrogen adsorption
are quite local. As can be seen in panels (b) - (e), the
H-saturated dimers are parallel to the (001) surface. 
The respective Si--Si dimer bond lengths are 1.6\% -- 2.6\%
larger than the Si--Si bulk bond length. The Si-H bond length 
amounts to 1.51 \AA.

\begin{table}[hbt]
\begin{center}
\begin{tabular}{clc}
     & adsorption site            & chemisorption energy [eV] \\
\hline
 (a) & H at rebonded S$_{\rm B}$  & -2.10  (-2.13) \\
 (b) & H on (1$\times$2) terrace  & -1.78  (-1.85) \\
 (c) & H on (1$\times$2) terrace  & -1.96  (-2.01) \\
 (d) & H on (2$\times$1) terrace  & -1.90  (-1.91) \\ 
 (e) & H on (2$\times$1) terrace  & -1.89  (-1.89) \\ 
\end{tabular}
\end{center}
\caption{
        Chemisorption energy for two H atoms adsorbed on a Si(1~1~11)
        surface with single atomic height steps. 
	The respective adsorption geometries are 
        shown in Fig.~\ref{fig:h2onsasb}. 
        The energies refer to a free H$_2$ molecule at rest
        far away from an S$_{\rm A}$-S$_{\rm B}$ vicinal surface,
        the ZPE correction is not included. 
	The data in parenthesis have been calculated at 30 Ry 
        cut-off energy with one special {\bf k} point. 
	The final energies have been calculated 
	at 50 Ry, using two special {\bf k} points in the
	irreducible part of the Brillouin zone, 
	and taking the frozen geometries from the 
	30 Ry, one {\bf k}-point run. 
        \label{tab:h2onsasb}
        }
\end{table}

The various chemisorption energies are summarized in Tab.~\ref{tab:h2onsasb}.
We find the adsorption at the rebonded S$_{\rm B}$ step-edge atom
to be energetically favored, with an energy difference
of at least 0.14 eV with respect to adsorption on terrace sites.
Note that for a comparison of the total energies of the final state 
configurations after hydrogen adsorption on, e.g., 
the S$_{\rm A}$ - S$_{\rm B}$ 
and the D$_{\rm B}$ stepped surface, the step energy difference 
of 0.02eV per step-length $2a$ has to be added to the numbers
given in Tab.~\ref{tab:h2onsasb}.

\subsection{Hydrogen adsorbed on surfaces with D$_{\rm B}$ steps}
\label{sec:rebdb}

\begin{figure}[htb]
  \epsfxsize=8.7cm
  \epsfbox{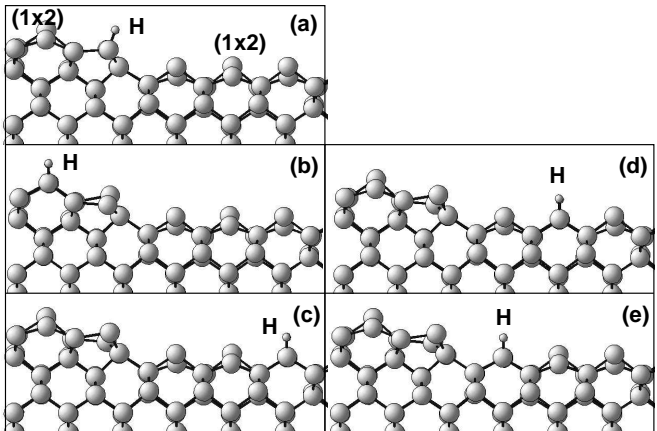}
  \caption{Relaxed atomic geometries for two hydrogen atoms adsorbed at
           different sites on a Si(1~1~11) surface with
           D$_{\rm B}$ steps. 
	   Side view along the [1\=10] direction as in 
           Fig.~\ref{fig:clean_surfaces}. 
           The hydrogen atoms are arranged in a row 
           parallel to [1\=10], thus only the atom in the front 
	   is visible.
	   (a) Two hydrogen atoms adsorbed at the D$_{\rm B}$ 
           step edge in a local mono-hydride configuration.
	   (b) -- (e) Two hydrogen atoms adsorbed at the 
           various inequivalent Si dimers on the 
	   (1$\times$2) terrace.
          }
  \label{fig:h2ondb}
\end{figure}

The relaxed geometries for two hydrogen atoms adsorbed on a 
vicinal silicon surface with D$_{\rm B}$ steps are shown in 
Fig.~\ref{fig:h2ondb}. The observed relaxations as well as the
chemisorption energies summarized in Tab.~\ref{tab:h2ondb} 
are similar to the results we found for the single
atomic height steps.

In Fig.~\ref{fig:h2ondb}(a) the two hydrogen atoms saturate the
dangling bonds of the two rebonded Si atoms at the step edge.
Consequently, the buckling of the step-edge atoms disappears.
The strain of the bond to the neighboring Si atom on the upper
terrace amounts to 5.3\% elongation, close to the average 
strain of these bonds on the clean surface. The buckling
of the surface dimers on the (1$\times$2) terrace remains almost
unaffected by hydrogen adsorption at the step edge.
Hydrogen adsorption on the terrace (Fig.~\ref{fig:h2ondb}(b)-(e)) 
leads to local mono-hydride configurations.
The H-saturated Si-dimer de-buckles, and the
Si--Si dimer bond-length expands by 1.9\% -- 2.5\%.

\begin{table}[hbt]
\begin{center}
\begin{tabular}{clc}
     & adsorption site            & chemisorption energy [eV] \\
\hline
 (a) & H at rebonded D$_{\rm B}$          & -2.09  (-2.11) \\
 (b) & H on (1$\times$2) terrace, pos. 1  & -1.77  (-1.82) \\
 (c) & H on (1$\times$2) terrace, pos. 2  & -1.92  (-1.95) \\
 (d) & H on (1$\times$2) terrace, pos. 3  & -1.97  (-2.00) \\ 
 (e) & H on (1$\times$2) terrace, pos. 4  & -1.97  (-2.01) \\ 
\end{tabular}
\end{center}
\caption{
        Chemisorption energy for two H atoms adsorbed on a Si(1~1~11)
        surface with double atomic height D$_{\rm B}$ steps. 
	The respective adsorption geometries are 
        shown in Fig.~\ref{fig:h2ondb}. 
        The energies refer to a free H$_2$ molecule at rest
        far away from a clean D$_{\rm B}$-stepped vicinal surface,
        the ZPE correction is not included. 
	The data in parenthesis have been calculated at 30 Ry 
        cut-off energy with one special {\bf k} point. 
	The final energies have been calculated 
	at 50 Ry, using two special {\bf k} points in the
	irreducible part of the Brillouin zone, 
	and taking the frozen geometries from the 
	30 Ry, one {\bf k}-point run. 
        \label{tab:h2ondb}
        }
\end{table}

\begin{table}[hbt]
\begin{center}
\begin{tabular}{lc}
adsorption site            & chemisorption energy [eV] \\
\hline
H at rebonded D$_{\rm B}$          & -2.11   \\
H on (1$\times$2) terrace, pos. 1  & -1.80   \\
H on (1$\times$2) terrace, pos. 4  & -1.97   \\ 
\end{tabular}
\end{center}
\caption{
        Chemisorption energy for two H atoms adsorbed on a Si(117)
        surface with double atomic height D$_{\rm B}$ steps. 
        The energies refer to a free H$_2$ molecule at rest
        far away from a clean D$_{\rm B}$-stepped Si(117) surface,
        the ZPE correction is not included. 
	The data have been calculated at 30 Ry 
        cut-off energy with two special {\bf k} points. 
        \label{tab:h2ondb_117}
        }
\end{table}

From the chemisorption energies in Tab.~\ref{tab:h2ondb} we read
that the H atoms preferentially adsorb on the rebonded Si atom at
the D$_{\rm B}$ step edge. In comparison to the adsorption sites
on the terrace the step is energetically favorable by at least 0.12 eV.
We speculate that this may be due to both the different elastic
relaxation energies at the step edge and on the terrace and some
residual $\pi$ bond between the surface dimer atoms as opposed 
to the step edge atoms. A rigorous quantitative analysis of 
the energy contributed by the different possible mechanisms, 
however, is beyond the scope 
of this paper.

Comparing the total energies for hydrogen attached to 
the D$_{\rm B}$ and S$_{\rm A}$ - S$_{\rm B}$ step edge we find
no significant energy difference within the accuracy of our 
approach. 

We have repeated our calculations for a smaller supercell 
describing a Si(117) surface with D$_{\rm B}$ steps, which 
has two Si dimers per (1$\times$2) terrace.
The chemisorption energies are summarized in Tab.~\ref{tab:h2ondb_117}.
They compare very well with the values for the Si(1~1~11) surface
in Tab.~\ref{tab:h2ondb}, i.e., we do not find any pronounced 
effect of miscut angle on chemisorption energies.
This lends support to the use of the smaller, and
thus computationally more convenient,  Si(117) supercell
for the calculation of, e.g., the hydrogen adsorption barriers
on vicinal Si(001) surfaces. 

A further issue that needs to be investigated is the existence
of different surface geometries corresponding to local minima
of the total energy in configuration space. 
For this purpose we have taken the configuration displayed in 
Fig.~\ref{fig:h2ondb}(d) and flipped the buckling angle
of all dimers to the left hand side or to the right hand side
of the mono-hydride group up  to the neighboring step edge
from $+\Theta$ to $-\Theta$, and vice versa. Then these
starting configurations were relaxed in the standard way.
In this way we found new stable configurations, which,
however, are all energetically degenerate within the accuracy
of our approach. 
At room temperature and above, the dimers 
flip rapidly between the two stable orientations.

The large strain of the bond between the D$_{\rm B}$ step-edge atom
(i.e., the rebonded Si atom) and the neighboring Si atom on the upper
terrace supports the speculation that this bond could break upon
hydrogen adsorption and that a local di-hydride group might form
at the position of the former step edge-atom.  
Therefore we have calculated the total energy
of the relaxed di-hydride geometry on a Si(117) surface, however,
we found it to be energetically
distinctly unfavorable. The respective hydrogen chemisorption energy 
amounts to only -0.92 eV.

\subsection{Hydrogen adsorbed on surfaces with non-rebonded 
            D$_{\rm B}^\prime$ steps}
\label{sec:nonrebdb}

\begin{figure}[htb]
  \epsfxsize=8.7cm
  \epsfbox{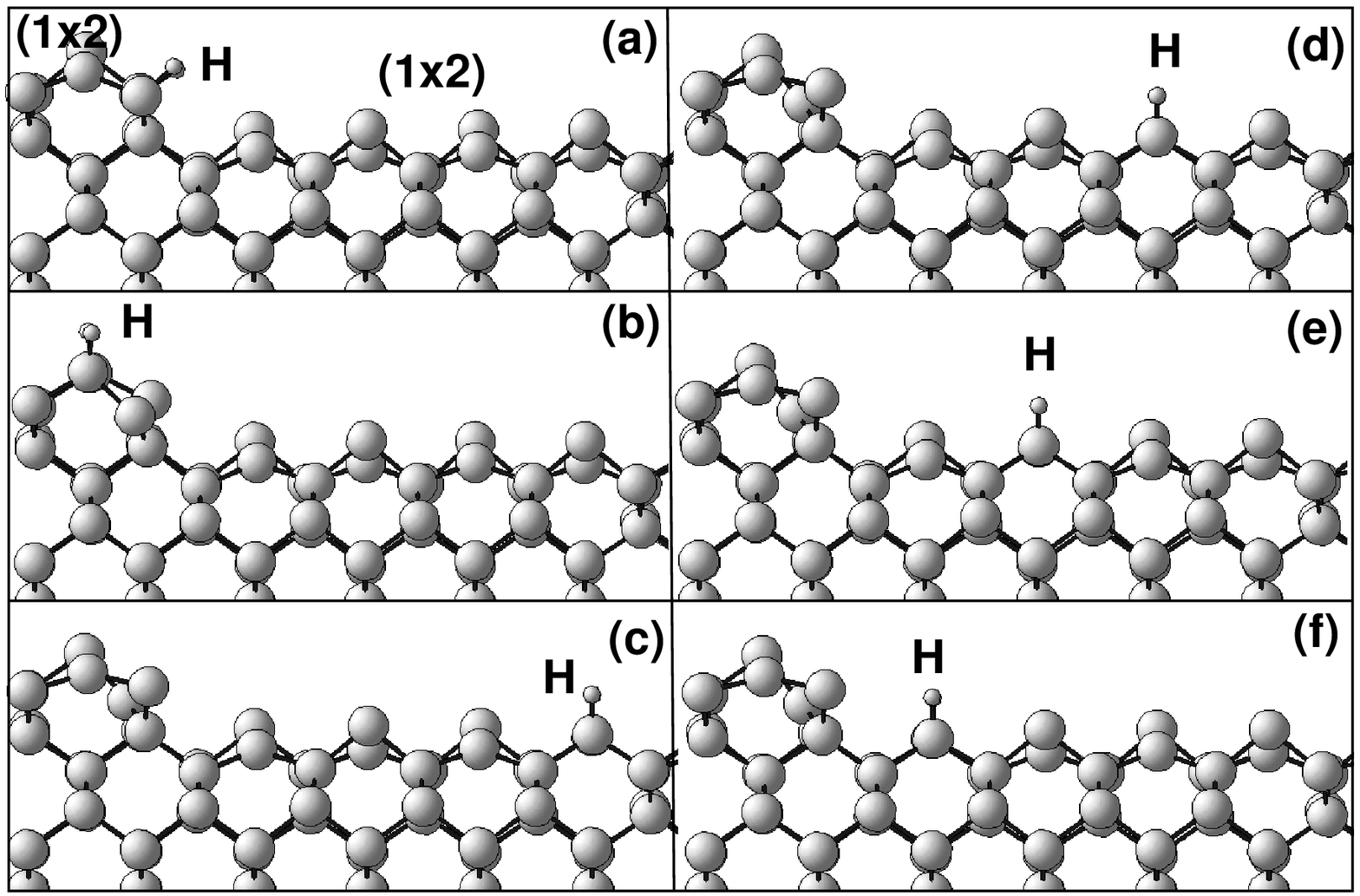}
  \caption{Relaxed atomic geometries for two hydrogen atoms adsorbed at
           different sites on a Si(1~1~11) surface with
           non-rebonded D$_{\rm B}^\prime$ steps. 
	   Side view along the [1\=10] direction as in 
           Fig.~\ref{fig:clean_surfaces}. 
           The hydrogen atoms are arranged in a row 
           parallel to [1\=10], thus only the atom in the front 
	   is visible.
	   (a) Two hydrogen atoms adsorbed at the D$_{\rm B}^\prime$ 
           step edge in a local mono-hydride configuration.
	   (b) -- (f) Two hydrogen atoms adsorbed on the 
           various inequivalent Si dimers on the 
	   (1$\times$2) terrace.
          }
  \label{fig:h2onnonrebdb}
\end{figure}

Relaxed geometries for two hydrogen atoms adsorbed on a 
Si(1~1~11) surface with non-rebonded D$_{\rm B}^\prime$ steps
are displayed in Fig.~\ref{fig:h2onnonrebdb}, and the respective
chemisorption energies are given in Tab.~\ref{tab:h2onnonrebdb}.
Interestingly, the non-rebonded step behaves totally different 
from the rebonded steps discussed in the previous Sections.
In particular, the hydrogen adsorption at the step 
edge (i.e., on the adsorption sites
shown in panels (a) and (b) of Fig.~\ref{fig:h2onnonrebdb})
is distinctly disfavored. We attribute this to the
breaking of the strong $\pi$ bonds between the step edge Si
atom and the neighboring Si surface-dimer atom on the upper 
terrace. This bond is contracted by 3\% on the clean surface
and in the configurations (c) - (f), while this contraction 
disappears upon hydrogen adsorption on either of the two sites 
shown in Fig.~\ref{fig:h2onnonrebdb}(a) and (b).

\begin{table}[hbt]
\begin{center}
\begin{tabular}{clc}
     & adsorption site            & chemisorption energy [eV] \\
\hline
 (a) & H at non-rebonded D$_{\rm B}^\prime$  & -1.53  \\
 (b) & H on (1$\times$2) terrace, pos. 1     & -1.27  \\
 (c) & H on (1$\times$2) terrace, pos. 2     & -1.93  \\
 (d) & H on (1$\times$2) terrace, pos. 3     & -1.93  \\ 
 (e) & H on (1$\times$2) terrace, pos. 4     & -1.93  \\ 
 (f) & H on (1$\times$2) terrace, pos. 5     & -2.02  \\ 
\end{tabular}
\end{center}
\caption{
        Chemisorption energy for two H atoms adsorbed on a Si(1~1~11)
        surface with non-rebonded D$_{\rm B}^\prime$ steps. 
	The respective adsorption geometries are 
        shown in Fig.~\ref{fig:h2onnonrebdb}. 
        The energies refer to a free H$_2$ molecule at rest
        far away from a vicinal Si surface with non-rebonded
	D$_{\rm B}^\prime$ steps.
        The ZPE correction is not included. 
	The calculations have been carried out 
	at 30 Ry, with one special {\bf k} point.
        \label{tab:h2onnonrebdb}
        }
\end{table}

On the whole, the absolute values of the chemisorption energies are 
smaller on the D$_{\rm B}^\prime$ stepped surface than on
the D$_{\rm B}$ stepped surface. Together with the fact that
already the clean D$_{\rm B}^\prime$ surface is energetically
unfavorable we conclude, that the D$_{\rm B}^\prime$ step
configuration is unstable also after hydrogen adsorption.
This holds true unless the hydrogen coverage becomes so large
that the hydrogen atoms cannot be accommodated anymore on the
D$_{\rm B}$ stepped surface. 
Our conclusion is that after adsorption of two hydrogen atoms
on a vicinal Si surface the most stable, i.e. lowest energy, 
configurations consist of the two H atoms
bound to either the rebonded S$_{\rm B}$ or the rebonded
D$_{\rm B}$ step edge.

\subsection{Partially hydrogen-covered surfaces}
\label{sec:hcov}

Next we consider finite hydrogen coverage and interaction effects 
between neighboring mono-hydride H--Si--Si--H groups on the surface.
This will allow us to investigate a possible tendency towards hydrogen 
island formation, and the energetical competition of such islands, 
acting as hydrogen sinks, with the step-edge adsorption-sites 
discussed in the previous Sections. 
To gain further principal insight into 
the energetics of hydrogen adsorption, we consider both the
rebonded $D_{\rm B}$ and the non-rebonded $D_{\rm B}^\prime$ step.
In this Section we still assume the limit of large pairing energy,
i.e., the H atoms are assumed to always form complete H--Si--Si--H 
groups.

For the description of our approach we will restrict ourselves to the
rebonded D$_{\rm B}$ step. The different adsorption sites are
enumerated from 0 to 4 as in Tab.~\ref{tab:h2ondb}, and the index
0 denotes the two rebonded Si atoms at the step edge. The H$_2$ 
chemisorption energies $\epsilon_i$ are taken from the (30 Ry cut-off energy, 
1 {\bf k} point) results in Tab.~\ref{tab:h2ondb}. This has been done for
the sake of a simple comparison to the results for the D$_{\rm B}^\prime$
step, for which only (30 Ry cut-off energy, 1 {\bf k} point) results 
are available,
and for the sake of less expensive {\it ab initio} calculations.
The numbers $n_i, i=0,...,4$, can take the values 0 or 1, with 0
denoting a clean Si surface dimer, and 1 denoting a Si surface dimer with
two adsorbed H atoms. We assume the following nearest-neighbor-interaction
Hamiltonian to describe the total chemisorption energy 
(take index 5 to be identical to index 0):
\begin{equation}
H^{({\rm D}_{\rm B})} = 
    \sum_{i=0}^{4} n_i \epsilon_i^{({\rm D}_{\rm B})} + 
    n_i n_{i+1} w_i^{({\rm D}_{\rm B})}. 
\label{model1}
\end{equation}
As a further approximation we assume the interaction parameters 
$w_i$ to be equal on the terrace, i.e., 
$w_1{({\rm D}_{\rm B})} = w_2{({\rm D}_{\rm B})} = w_3{({\rm D}_{\rm B})}$.
Effects omitted in this Hamiltonian are ({\it i}) the breaking-up
of H--Si--Si--H pairs into single adsorbed H atoms, ({\it ii}) the
excitation of the D$_{\rm B}$ step into S$_{\rm A}$ - S$_{\rm B}$
step pairs, and ({\it iii}) any effect connected with the 
configurational entropy due to dimer flips.
The three independent interaction parameters $w_i$ have been determined 
within DFT (at 30 Ry cut-off energy, with 1 {\bf k} point), 
by calculating two geometries with four adsorbed H atoms and 
the fully H-covered surface. 
The results for the rebonded D$_{\rm B}$ step are:
$w_0{({\rm D}_{\rm B})} = -0.14$eV,
$w_1{({\rm D}_{\rm B})} = -0.04$eV,
$w_4{({\rm D}_{\rm B})} = -0.009$eV,
and similar calculations for the non-rebonded D$_{\rm B}^\prime$ 
step yield:
$w_0{({\rm D}_{\rm B}^\prime)} = -0.75$eV,
$w_1{({\rm D}_{\rm B}^\prime)} = -0.07$eV,
$w_5{({\rm D}_{\rm B}^\prime)} = -0.007$eV.
Obviously, the nearest-neighbor interaction is attractive.
We attribute it mostly to elastic interaction via the second and deeper
layer Si atoms. The extraordinarily large value of 
$|w_0{({\rm D}_{\rm B}^\prime)}|$ accounts for the fact that 
two Si--Si $\pi$ bonds have already been broken by the first H-pair
(configurations Fig.~\ref{fig:h2onnonrebdb}(a) and (b)),
thereby favoring the adsorption of the second H-atom pair 
on the neighboring Si atoms.

Interesting quantities that can easily be calculated 
from Eq.~(\ref{model1}) are the chemisorption ``hole'' energies,
i.e., the chemisorption energy of the last H$_2$ molecule
that is finally completing the full coverage of one monolayer.
We label the adsorption sites in analogy to Figs.~\ref{fig:h2ondb} and
\ref{fig:h2onnonrebdb}; to obtain the initial configuration before
adsorption just invert the surface H-occupation (i.e., every Si
dangling bond becomes H-covered, apart from those two Si 
dangling bonds that are saturated by H atoms in the respective 
panel of Figs.~\ref{fig:h2ondb} or \ref{fig:h2onnonrebdb}). 
The chemisorption ``hole'' energies are summarized in 
Tabs.~\ref{tab:holedb} and \ref{tab:holenonrebdb} for
D$_{\rm B}$ and D$_{\rm B}^\prime$ stepped surfaces, respectively.
The accuracy of our model Hamiltonian is demonstrated by
the good agreement with two DFT test calculations, 
in which the atomic geometries have been fully relaxed.
Contrary to adsorption on the clean surfaces studied above,
we find the rebonded and non-rebonded step configurations to
behave very similar in case of the ``hole'' 
energies. In particular, for both steps the absolute value of the 
chemisorption energy takes its maximum for adsorption at
the step edge. Furthermore, the chemisorption energies on the
terraces in Tabs.~\ref{tab:holedb} and \ref{tab:holenonrebdb}
are all quite similar and below (i.e., more negative than) 
the values for the clean surfaces.
This simplicity is due to two reasons: First, the $\pi$ bond
breaking at the D$_{\rm B}^\prime$ step does not play a role
for the hole energies. Second, there is an elastic coupling 
between the relaxation of the buckled dimers on the surface 
and the step edge atoms on the clean surface, which apparently 
becomes less important when the Si atoms are H-covered and
the surface Si atoms are not buckled any more. 
The hydrogen adsorption into the ``hole'' configuration is
more exothermic than the adsorption on the clean surface
due to the different elastic interactions:  
The dimer buckling on the clean surface tends to stabilize
the buckled dimer and thus to de-stabilize the symmetric dimer of
the local mono-hydride group. 
Our results are compatible with the energy difference of 0.05 eV/dimer
between the asymmetric p(2$\times$1) and the p(2$\times$2)
reconstruction of the Si(001) surface calculated by
Ramstad {\it et al.}\cite{ramstad:95}

\begin{table}[htb]
\begin{center}
\begin{tabular}{cll}
     & adsorption site                & chemisorption energy [eV] \\
\hline
 (a) & rebonded D$_{\rm B}$ step edge & \hspace{1cm} -2.26 (-2.25) \\ 
 (b) & (1$\times$2) terrace, pos. 1   & \hspace{1cm} -2.00 \\
 (c) & (1$\times$2) terrace, pos. 2   & \hspace{1cm} -2.03 \\
 (d) & (1$\times$2) terrace, pos. 3   & \hspace{1cm} -2.08 (-2.07) \\
 (e) & (1$\times$2) terrace, pos. 4   & \hspace{1cm} -2.06 \\
\end{tabular}
\end{center}
\caption{ 
	Chemisorption ``hole'' energy for an almost fully 
        hydrogen-covered Si(1~1~11) surface with D$_{\rm B}$ steps.
	The energies have been derived from the model Hamiltonian
	in Eq.~(\ref{model1}). For comparison, the energy values in 
	parenthesis are from {\it ab initio} total-energy calculations
        (30 Ry cut-off energy, 1 {\bf k} point). For the 
	meaning of the labels denoting the position of the hole
	see Fig.~\ref{fig:h2ondb}.
        \label{tab:holedb}
        }
\end{table}

\begin{table}[hbt]
\begin{center}
\begin{tabular}{clc}
     & adsorption site            & chemisorption energy [eV] \\
\hline
  (a) & non-rebonded D$_{\rm B}^\prime$ step edge & -2.28 \\
  (b) & (1$\times$2) terrace, pos. 1              & -2.09 \\
  (c) & (1$\times$2) terrace, pos. 2              & -2.07 \\
  (d) & (1$\times$2) terrace, pos. 3              & -2.07 \\
  (e) & (1$\times$2) terrace, pos. 4              & -2.07 \\
  (f) & (1$\times$2) terrace, pos. 5              & -2.10 \\
\end{tabular}
\end{center}
\caption{
	Chemisorption ``hole'' energy for an almost fully 
        hydrogen-covered Si(1~1~11) surface with 
	non-rebonded D$_{\rm B}^\prime$ steps.
	The energies have been derived from a model Hamiltonian
	similar to Eq.~(\ref{model1}). For the 
	meaning of the labels denoting the position of the hole
	see Fig.~\ref{fig:h2onnonrebdb}.
        \label{tab:holenonrebdb}
        }
\end{table}

\begin{figure}[htb]
  \epsfxsize=8.7cm 
  \centerline{\epsfbox{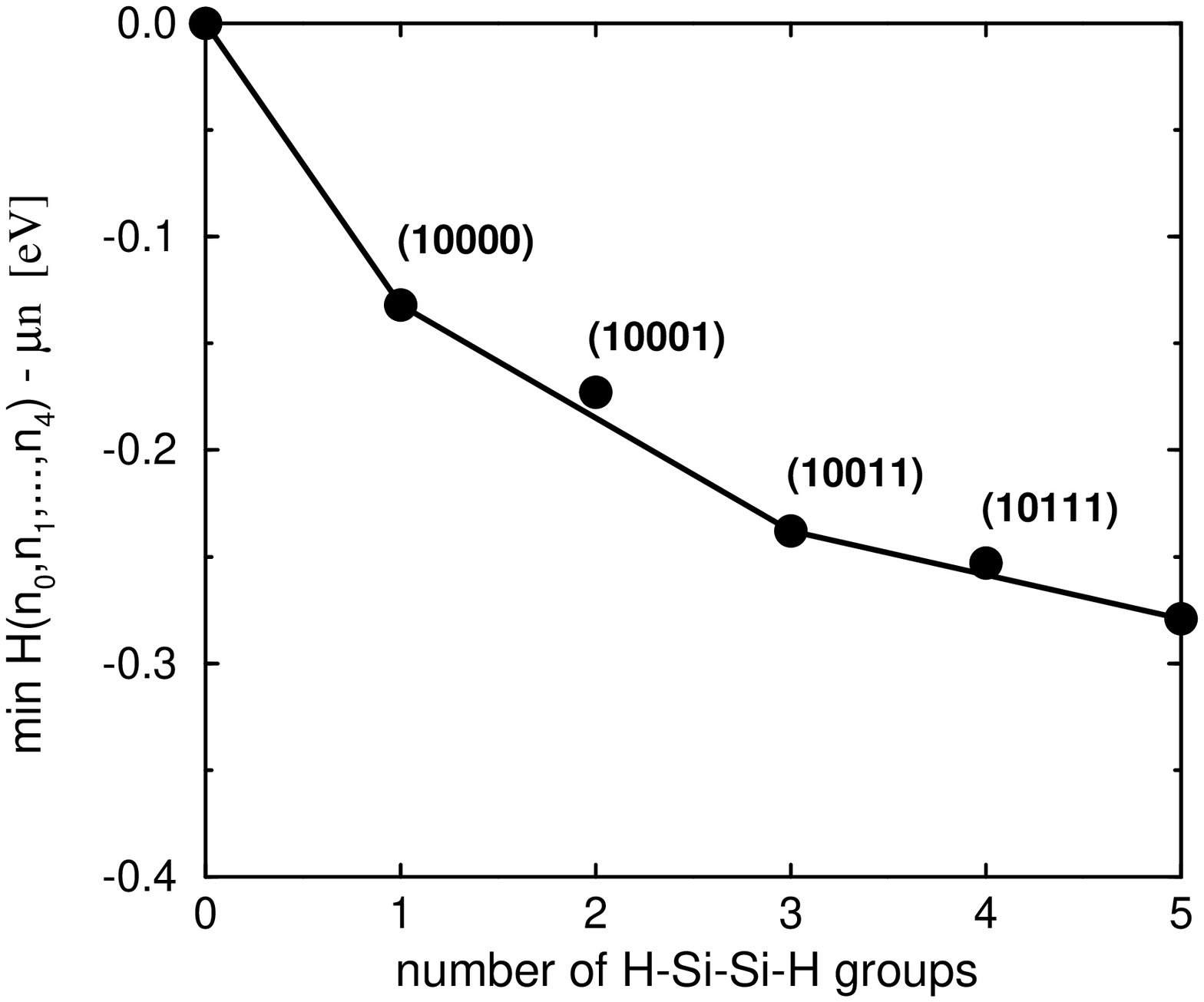}}
  \caption{
          The dots represent the minimum energy 
	  $\min( H(n_0, n_1, ..., n_4)) -\mu n$ with respect
	  to all configurations ($n_0, n_1, ..., n_4$) at fixed $n$
	  versus the number of H--Si--Si--H
	  groups $n=\sum_{i=0}^4 n_i$, for a Si(1~1~11) surface
	  with D$_{\rm B}$ steps.
	  The labels denote the lowest energy configurations.
	  A chemical potential term $\mu n$ has been subtracted for 
	  convenience, with $\mu$ chosen equal to the average single 
	  particle chemisorption energy of -1.98 eV.
	  The model Hamiltonian $H$ from Eq.~(\ref{model1}) is used 
	  together with the {\it ab initio} energy parameters calculated
	  with 1 {\bf k} point at 30 Ry cut-off energy. 
	  The full line is the convex hull, which represents the
	  thermodynamic ground state energy (for 
          temperature $T \rightarrow 0$K) as a function of $n$.
          }
  \label{fig:total_energies}
\end{figure}

Up to this point we have focused on the chemisorption energies 
as a function of
the different adsorption sites, and therefore it was sufficient to 
consider only a single 
dimer string. In the following we will shortly discuss 
the ``thermodynamic ground state'' of the whole surface as a function of
hydrogen coverage. This means that, while we still stick to the simple
Hamiltonian in Eq.~(\ref{model1}) to describe the energy of every 
single dimer row on the $D_{\rm B}$ stepped surface, 
we additionally allow for the exchange of hydrogen atom pairs 
between adsorption sites on different dimer rows. 
In this way chemisorption configurations with inhomogeneous hydrogen
coverage are included in the competition for the lowest total energy.
In Fig.~\ref{fig:total_energies} the minimum of the energy given by
Eq.~(\ref{model1}) with respect to all configurations $(n_0, n_1, ..., n_4)$,
$n_i = 0$ or 1, is plotted versus the number of H--Si--Si--H pairs
$n=n_0+ ... +n_4$. The full line denotes the convex hull and represents
the lowest possible energy as a function of $n$, also accounting for
the simultaneous occurrence of different chemisorption 
configurations on the surface.
Fig.~\ref{fig:total_energies} leads us to the important conclusion that 
the vicinal surface where all the dangling bonds of the rebonded Si atoms 
at the D$_{\rm B}$ step edge are saturated by H atoms (configuration
$(n_0, n_1, ..., n_4) = (1,0,0,0,0)$) in fact represents 
the ground state at this certain coverage (i.e., $\Theta=1/5$ for 
the Si(1~1~11) surface).
Obviously, the interaction included in Eq.~(\ref{model1}) is too 
weak to stabilize any other state with hydrogen ``islands'' 
on the terrace.

\begin{figure}[htb]
  \epsfxsize=7.0cm 
  \centerline{\epsfbox{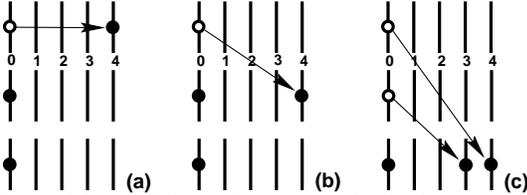}}
  \caption{
	Metastable configurations for hydrogen adsorbed on a
	Si(1~1~11) surface with D$_{\rm B}$ steps and hydrogen
	coverage 1/5. The Si
	surface dimers and the step-edge atoms are denoted by 
	lines. On the terrace the dimers arrange in dimer strings.
	The open and filled circles denote the position
	of the mono-hydride H--Si--Si--H groups in the ground
	state and in the excited state configuration, respectively.
	(a) One mono-hydride group displaced from the step edge
	onto the terrace within one dimer row.
	(b) One mono-hydride group displaced from the step edge
	onto a terrace site of another dimer row.
	(c) Two mono-hydride groups displaced onto another
	single dimer row.
          }
  \label{fig:excitations}
\end{figure}

In Fig.~\ref{fig:excitations} the lowest energy configurations are
shown for the two situations when one or two H--Si--Si--H groups are
displaced from the D$_{\rm B}$ step edge onto the terrace.
In panel (a) the hydrogen group is forced to remain on its dimer
string, thus the energy difference can be immediately read 
from Tab.~\ref{tab:h2ondb} to be equal to 0.10 eV 
(here we consistently use the results for 30 Ry, 1 {\bf k} point; 
for 50 Ry, 2 {\bf k} points the energy difference would amount to 0.12 eV). 
The energy difference becomes slightly smaller, 0.09 eV, for
the configuration displayed in panel (b), when the 
hydrogen atoms are transferred onto another dimer string.
Finally, when two hydrogen groups are displaced 
as shown in panel (c), the energy difference amounts to 0.16 eV.

\subsection{Thermodynamics at low hydrogen coverage}
\label{sec:lowcov}

\begin{figure}[htb]
  \epsfxsize=7.5cm 
  \centerline{\epsfbox{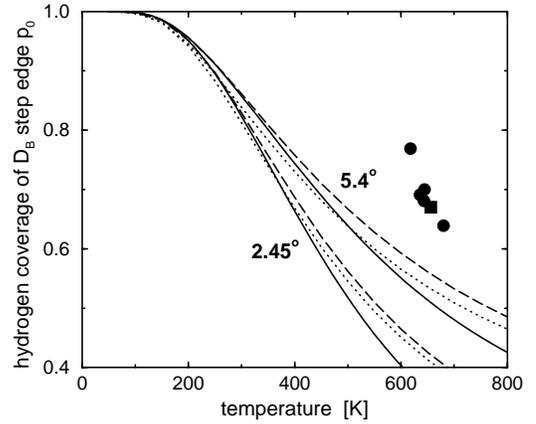}}
  \caption{
	Probability $p_0$, that the Si dangling bond at the D$_{\rm B}$ 
	step edge is saturated with a hydrogen atom, versus the
	surface temperature for two vicinal surfaces with miscut angles
	of 2.45$^\circ$ and 5.4$^\circ$ degrees.  
	The total hydrogen coverage amounts to two hydrogen atoms
	per dimer string, i.e., it has been chosen equal to $1/(N+1)$
	independent of temperature.
	Different line styles are used to distinguish among the
	various approximations. Dotted line: Hamiltonian Eq.~(\ref{model1}),
	including the nearest neighbor interaction term. 
	Dashed line: Same Hamiltonian, but all interactions 
	$w_i$ set to zero. Full line: Hamiltonian Eq.~(\ref{model2})
	with pairing energy $\epsilon_{\rm pair} = 0.25$eV for
	all sites.
	The symbols denote experimental data by Raschke and H\"ofer
	for vicinal surfaces with a miscut angle of 2.5$^\circ$ (circles)
	and 5.5$^\circ$ (square).
	  }
  \label{fig:p0}
\end{figure}

\begin{figure}[htb]
  \epsfxsize=7.5cm 
  \centerline{\epsfbox{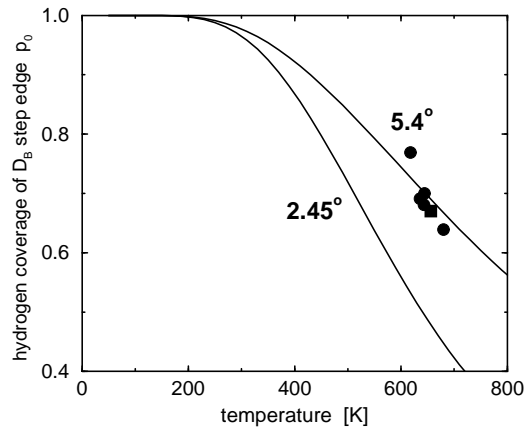}}
  \caption{
  	Same as Fig.~\ref{fig:p0}. However, in this case only 
  	partial thermal equilibration of the system is assumed:
	The adsorbed hydrogen atoms are allowed to diffuse 
	only along the dimer row, or to hop from one Si atom
	to the other Si atom within the same dimer. 
	Results are based on the Hamiltonian in Eq.~(\ref{model2})
	and a canonical ensemble.
	The symbols denote experimental data by Raschke and H\"ofer
	for vicinal surfaces with a miscut angle of 2.5$^\circ$ (circles)
	and 5.5$^\circ$ (square).
	  }
  \label{fig:p0_canonical}
\end{figure}

In this Section we will present a comparison between our {\it ab initio}
results and the experimental hydrogen adsorption data by 
Raschke and H\"ofer described in the preceding paper.\cite{raschke:98c}
They have saturated the D$_{\rm B}$ step edges of vicinal Si(001) 
surfaces miscut by 2.5$^\circ$ and 5.5$^\circ$ towards the [110] direction
with molecular hydrogen at a surface temperature, 
which is sufficiently low to 
suppress the diffusion of the hydrogen atoms onto the terrace.  
This non-equilibrium configuration can be prepared, because
there is a considerable energy barrier towards the dissociative
adsorption of the H$_2$ molecules onto the flat surface or 
the terrace, while the adsorption-energy barrier at the step edge is very 
small.\cite{raschke:98a,raschke:98b}
At temperatures between 618 K -- 680 K partial thermal equilibrium 
between the adsorption sites at the step and on the terrace
is established. After equilibration the configuration is frozen-in 
by rapidly cooling down the sample, and the residual  
hydrogen coverage at the D$_{\rm B}$ step-edge site is 
measured. This is accomplished by determining the
amount of hydrogen, expressed as a change of coverage $\Delta\Theta$, 
necessary to saturate the depleted Si step-edge dangling-bonds again. 
The resulting step-edge H-occupation-probability
$p_0 = 1 - \Delta\Theta/\Theta_{\rm step}^{\rm sat}$, based on the
experimental step-edge H-saturation coverage 
$\Theta_{\rm step}^{\rm sat}$, is denoted by the filled circles and 
the square in Figs.~\ref{fig:p0} and \ref{fig:p0_canonical}. 

To compare to experiment we generalize the Hamiltonian in Eq.~(\ref{model1})
from the Si(1~1~11) surface with $N$=4 Si dimers per terrace to 
D$_{\rm B}$-stepped vicinal surfaces with smaller miscut angle
and $N$$>$4 dimers per terrace. This is done by simply replicating the
energy corresponding to configuration (c) on the D$_{\rm B}$ stepped
surface (see Fig.~\ref{fig:h2ondb}) together with the interaction 
parameter $w_i$(D$_{\rm B}$), which is assumed to be constant on
the terrace. 
The configuration (c) was chosen, because its chemisorption energy 
is closest to the chemisorption energy on the terraces bounded
by D$_{\rm B}^\prime$ steps. As already noted above, the 
strain field of the non-rebonded 
D$_{\rm B}^\prime$ step is distinctly weaker than that of the 
rebonded D$_{\rm B}$ step. Therefore the terraces on the D$_{\rm B}^\prime$
stepped surface resemble the flat surface (or, equivalently,
the middle part of a wide terrace) more closely than any of the
other configurations.
For the miscut angles 5.4$^\circ$ and 2.45$^\circ$, corresponding to   
$N$=6 and $N$=15 dimers per terrace,
we have calculated the step-edge occupation probability 
$p_0 = \langle n_0 \rangle $ 
at fixed hydrogen coverage $\Theta = 1/(N+1)$
as a function of surface temperature within the grand canonical ensemble,
see the dotted line in Fig.~\ref{fig:p0}. 
At this small hydrogen coverage the interaction term in the Hamiltonian
(\ref{model1}) does not play any important role, as can be judged 
by comparison to the dashed line in Fig.~\ref{fig:p0}, which was 
calculated in exactly the same way, but with all interaction parameters
$w_i$ set to zero.

However, it is well known\cite{hoefer:92} that at small hydrogen coverage
configurational entropy dominates over the hydrogen pairing energy.
Single, un-paired chemisorbed H-atoms become more and more frequent 
at low H-coverage. 
To account for this process we generalize our model Hamiltonian.
The occupation numbers $n_i^A$ and $n_i^B$ can take the values 0 or 1,
depending on whether the Si atom ``$A$'' or ``$B$'' of the $i$-th
Si surface-dimer is H-saturated or not.  
For simplicity we omit the interaction between H-atoms on different 
dimers, i.e., we set all $w_i$ to zero.
Furthermore we assume a site-independent pairing energy $\epsilon_{\rm pair}$
of 0.25 eV, in agreement with previous 
experimental and theoretical work for the flat 
surface.\cite{hoefer:92,northrup:95}
The Hamiltonian is:

\begin{eqnarray}
H^{({\rm D}_{\rm B})} & = &
    \sum_{i=0}^{N} \{{1 \over 2}(n_i^A + n_i^B) \epsilon_i^{({\rm D}_{\rm B})} +
    \nonumber\\
 &  & +{1 \over 2} (n_i^A (1-n_i^B) + n_i^B (1-n_i^A)) \epsilon_{\rm pair}\} . 
\label{model2}
\end{eqnarray}

We have calculated the H-occupation probability at the D$_{\rm B}$ step edge,
$p_0 = \langle n_0^A + n_0^B \rangle / 2 $ for this model. The
results are plotted as full lines in Fig.~\ref{fig:p0}.
Among the model Hamiltonians discussed above we expect this approach
to yield the most realistic description of the Si surface in
thermodynamic equilibrium at low H-coverage.
Theory and experiment agree with respect to the fact that the
hydrogen more tightly binds to the step edge than on the terrace.
However, as can be read from Fig.~\ref{fig:p0}, there remains a
slight quantitative discrepancy.
Because the chemisorption energies at the
D$_{\rm B}$ and at the S$_{\rm B}$ step edge are similar,
we do not attribute this
difference to the existence of single atomic height steps on
the experimental sample.
Instead, we argue that at least part of the difference may be due to the
incomplete thermal equilibration of the surface in experiment. 
There is qualitative agreement between 
theory\cite{vittadini:93,carter:94,nachtigall:95}
and experiment\cite{owen:96} that the diffusion of hydrogen atoms
on the Si(001) surface is highly anisotropic and fast along the 
dimer rows. The experimental\cite{owen:96} activation energy 
for hopping along a dimer row amounts to $E_B$=1.68 eV 
when an attempt frequency of $\nu_0$=10$^{13}$ s$^{-1}$ is 
assumed. At a surface temperature of $T$=620 K this corresponds 
to a hopping rate $\nu=\nu_0 \exp(-E_B/k_BT)=$0.2 s$^{-1}$, i.e., 
many of these diffusion events occur during the time interval
of the order of 10$^3$ s during which the
surface is kept at elevated temperature in H\"ofer and 
Raschke's experiment. Similarly, the intra-dimer hopping 
is predicted to be as fast as the intra-row hopping
of the hydrogen atoms.\cite{vittadini:93} 
Thus we may safely assume a perfect thermal
scrambling of the H-atom adsorption sites within a single dimer row.
The hydrogen inter-row diffusion, on the other hand, is distinctly slower.
The theoretical barriers vary from 1.8 eV\cite{vittadini:93},
which translates into a hopping rate of roughly 0.02 s$^{-1}$, 
to barriers much larger than 2 eV\cite{carter:94,nachtigall:95},
where hopping perpendicular to the dimer rows would be 
almost completely suppressed. 
To investigate this issue we have calculated the thermal expectation
value $\langle n_0^A + n_0^B \rangle / 2$ for the 
hydrogen coverage at the step
edge from the Hamiltonian (\ref{model2}) within the canonical
ensemble. In this case the number of 
hydrogen atoms on every single dimer string is restricted 
to be equal to 2.  The result in Fig.~\ref{fig:p0_canonical}
shows a better agreement with Raschke and H\"ofer's experiment than 
Fig.~\ref{fig:p0}, 
i.e., our calculations are consistent with the notion that
at 620 K -- 680 K the hopping of H atoms across the
row and the exchange of H atoms between rows on different terraces
is indeed a rare process 
within a typical experimental time interval of 10$^2$ -- 10$^3$ s.

\section{Summary and conclusions}
\label{sec:summary}

The chemisorption energies for H$_2$ on vicinal Si(001) surfaces
have been calculated using a well established {\it ab initio} 
total-energy technique.\cite{fhi96md} 
The various adsorption sites on the terraces and at 
the step edges have been compared to each other. 
Both the single atomic height S$_{\rm A}$--S$_{\rm B}$
steps and the rebonded and non-rebonded double 
atomic height D$_{\rm B}$ and D$_{\rm B}^\prime$ steps have been 
considered. 
The results confirm recent experimental findings
by Raschke and H\"ofer\cite{raschke:98c} for the adsorption
of molecular hydrogen on vicinal Si surfaces.

In particular we conclude from our calculations:
\\
({\it i}) On a surface with D$_{\rm B}$ steps hydrogen preferentially 
binds to the step-edge Si atoms. The chemisorption energy difference 
with respect to the terrace sites is $\ge 0.12$ eV per H$_2$ molecule. 
\\
({\it ii}) Adsorption on a vicinal surface with single atomic height 
steps is similar to adsorption on the D$_{\rm B}$ stepped surface.
The hydrogen atoms preferentially bind to the S$_{\rm B}$ step edge,
with nearly the same chemisorption energy as for the D$_{\rm B}$
step edge.
\\ 
({\it iii}) The non-rebonded D$_{\rm B}^\prime$ step,
which is known to be unstable on the clean 
surface,\cite{chadi:87,oshiyama:95} remains 
unstable also after hydrogen adsorption. 
\\ 
({\it iv}) The interaction between neighboring mono-hydride
groups on the Si surface has been calculated and was found
to be locally attractive. However, this interaction is too weak to 
be of importance for the conclusions drawn in this paper.
\\
({\it v}) The hydrogen coverage of the D$_{\rm B}$ 
step-edge has been calculated as function of temperature. 
We assume that the diffusing hydrogen atoms are confined to a
single dimer row and obtain
qualitative and even quantitative agreement 
with the experiment.\cite{raschke:98c}

This work supplies the basis for a theoretical 
investigation of the H$_2$ adsorption energy barriers on vicinal 
Si(001) surfaces, especially for a comparison between the dissociative 
adsorption at the step edge and on the Si dimers of the terrace.

\section*{Acknowledgment}
\label{sec:acknowledgement}

We are grateful to M.~Scheffler for his continuous support 
of this work and  thank him, W.~Brenig, U.~H\"ofer, and M.~Raschke 
for enlightening discussions.
This work was supported in part by the Sfb 338 of the Deutsche 
Forschungsgemeinschaft.


\bibliographystyle{prsty}


\end{document}